\documentclass[aps,twocolumn,reprint, nofootinbib,preprintnumbers,superscriptaddress,citeautoscript]{revtex4-1}
\pdfoutput=1

\usepackage[utf8]{inputenc}

\usepackage[dvips]{color}
\usepackage[normalem]{ulem}
\usepackage{amsmath}
\usepackage{amssymb}
\usepackage{amscd}
\usepackage{enumerate}
\usepackage{amsfonts}
\usepackage{epsfig}
\usepackage{yfonts}

\usepackage{dsfont}

\usepackage{graphicx}
\usepackage{bm}

\definecolor{davecolor}{rgb}{0.95,  0.5,  0.2}

\def\({\left(}
\def\){\right)}
\def\[{\left[}
\def\]{\right]}
\def\<{\langle}
\def\>{\rangle}

\newcommand{\myfig}[3]{
	\begin{figure}[ht]
	\centering
	\includegraphics[width=#2cm]{#1}\caption{#3}\label{fig:#1}
	\end{figure}
	}

\newcommand{\be}{\begin{equation}}
\newcommand{\ee}{\end{equation}}
\newcommand{\bea}{\begin{eqnarray}}
\newcommand{\eea}{\end{eqnarray}}
\newcommand{\bwt}{\begin{widetext}}
\newcommand{\ewt}{\end{widetext}}

\newcommand{\bi}{\begin{itemize}}
\newcommand{\ei}{\end{itemize}}
\newcommand{\ben}{\begin{enumerate}}
\newcommand{\een}{\end{enumerate}}
\newcommand{\bca}{\begin{cases}}
\newcommand{\eca}{\end{cases}}
\newcommand{\bln}{\begin{align}}
\newcommand{\eln}{\end{align}}
\newcommand{\bst}{\begin{split}}
\newcommand{\est}{\end{split}}

\begin{document}

\title{Gravitational dual of the R\'{e}nyi twist displacement operator}
\author{Srivatsan Balakrishnan, Souvik Dutta, Thomas Faulkner}
\address{Department of Physics, University of Illinois, 1110 W. Green St., 
Urbana IL 61801-3080, U.S.A}
\begin{abstract}
We give a recipe for computing correlation functions of the displacement operator localized on a spherical or planar higher dimensional twist defect using AdS/CFT.  
Such twist operators are typically used to construct the  $n$'th Renyi entropies of spatial entanglement in CFTs  and are holographically dual to black holes with hyperbolic horizons. The displacement operator then tells us how the Renyi entropies change under small shape deformations of the entangling surface.
We explicitly construct the bulk to boundary propagator for the displacement operator insertion
as a linearized metric fluctuation of the hyperbolic black hole and use this to 
extract the coefficient of the displacement operator two point function $C_D$ in any dimension.
The $n \rightarrow 1$ limit of the twist displacement operator gives the same bulk response as the insertion of a null energy operator in vacuum, which is  consistent with recent results on the shape dependence of entanglement entropy and modular energy. 
\end{abstract}

\maketitle

\noindent \emph{Introduction.} The study of entanglement  in quantum field theories (QFT) has led to new perspectives on strongly correlated phenomena. Some recent applications include insight into the renormalization group flow in the space of QFTs \cite{Casini:2004bw,Liu:2012eea,Casini:2012ei}, the dynamics of excited states \cite{Calabrese:2005in,Hartman:2013qma,Liu:2013iza}, topological phases \cite{Kitaev:2005dm,PhysRevLett.96.110405,PhysRevLett.101.010504}, emergent space-time and gravitational dynamics hidden behind holographic dualities \cite{VanRaamsdonk:2010pw,Lashkari:2013koa,Faulkner:2013ica} and proofs of energy conditions in QFTs \cite{Blanco:2013lea,Bousso:2015wca,Faulkner:2016mzt}.  Conformal field theories (CFTs) offer hope for studying entanglement in strongly correlated systems where the available symmetries fit nicely with entanglement computations \cite{holzhey1994geometric,Calabrese:2004eu,Casini:2011kv}.  CFTs live at quantum critical points where long range entanglement is a basic characteristic \cite{vidal2003entanglement,fradkin2006entanglement}. Indeed the spatial Renyi entropies, the entanglement measure we plan to study in this paper, can even be studied in the lab \cite{islam2015measuring} and are often employed in numerical modeling  \cite{hastings2010measuring}, providing signatures of quantum phase transitions and topological phases. All of the above motivates development of theoretical tools for the study of Renyi entropies in CFTs and in this paper we will add to already available holographic results \cite{hung2011holographic,headrick2010entanglement,Faulkner:2013yia,hartman2013entanglement,dong2016shape,camps2016gravity} to meet this goal.

The Renyi entanglement entropies between a spatial region $A$ and its complement  can be constructed via the replica trick which computes $ {\rm Tr} \rho^n_A$ for integer $n$ by considering
a product orbifold CFT$^n/\mathbb{Z}_n$ and introducing a co-dimension-2 cyclic twist defect on the boundary $\partial A$ separating the spatial regions. An analytic continuation away from integer $n$ for correlation functions involving twist defects is often assumed, although is sometimes subtle \cite{Faulkner:2014jva,Agon:2015ftl}.
To exploit the maximal symmetries of the problem it has been recently suggested \cite{Bianchi:2015liz} that we should use the full technology of defect CFTs (dCFTs). Indeed the theoretical study of CFTs, including dCFTs \cite{Gaiotto:2013nva,Billo:2013jda,Liendo:2012hy}, in higher than 2 dimensions has recently undergone a renaissance \cite{Poland:2016chs}. We might hope to exploit as many of these results as possible in the study of entanglement.  

In this paper we plan to study the displacement operator localized on the Renyi defect. This operator is fundamental to any dCFT and has the property of moving the defect locally. Thus its correlation functions constrain the shape dependence of the Renyi entropies, controlling such observables as the universal log divergences from scale anomalies \cite{Lee:2014xwa,Lewkowycz:2014jia} and corner terms \cite{Bueno:2015rda,Bueno:2015qya,Bueno:2015lza,Bueno:2015ofa}. As a basic object in dCFT, it is of interest to construct its holographic dual \cite{maldacena1999large}.  There are many examples of holographic duals for dCFTs \cite{DeWolfe:2001pq,D'Hoker:2007fq,D'Hoker:2007xy,Gomis:2008qa,harrison2012maximally,Dias:2013bwa} of various co-dimensions, the Renyi entropies \cite{hung2011holographic} are just one example.  To our knowledge no construction of the holographic displacement operator has previously appeared,  with the exception of dCFTs described by probe branes \cite{DeWolfe:2001pq} where the displacement operator decouples to leading order from the bulk CFT. 
The construction of holographic defect operators has previously been worked out in \cite{Aharony:2003qf}; however  the displacement operator was not considered. 
In this paper we plan to remedy this, at least for the Renyi twist defects. We expect our methods generalize to the many known holographic dCFTs. 

\vspace{8pt}
\noindent \emph{Displacement operator.} We start by summarizing the known dCFT constraints on the displacement operator correlation functions.
These were recently elucidated in  \cite{Billo:2016cpy} using the embedding space formalism \cite{Costa:2011mg}. We will present their results in a way which is convenient and suggestive for translating into holography. 

The displacement operator is defined via a variation of the location of the twist defect:
\be
\label{deform}
\delta_w \tau_\Sigma = - \int_\Sigma \delta w^\alpha D_\alpha \tau_\Sigma
\ee
where $D_\alpha$ is a local operator living on the defect.
It has scaling dimension $d-1$ and is a vector under $SO(2)$ rotations in the transverse plane. 

The Ward identity relates $D_\alpha$ to the divergence of the stress tensor in the presence of the twist
defect, which thus encodes the breaking of translation invariance:
\be
\label{wi}
\partial_\mu T^{\mu}_{\,\,\,\,\alpha} = D_\alpha  \delta_{\Sigma}  
\ee
where $\delta_\Sigma$ is a $2$-dimensional delta function in the transverse plane localized along $\Sigma$.

While these definitions work for arbitrary shaped twist defects, to make any further statements
we need to specialize to flat space with a planar defect.
Such a defect globally preserves an $SO(2) \times SO(d-1,1)$ subgroup of the flat space conformal group. The $SO(d-1,1)$ group acts as standard conformal transformations along the $d-2$ dimensional
defect. Note that while the detailed results will be presented for a planar defect, one can easily translate these results to spherically shaped defects.

The correlation function of two displacement operators along this planar defect is then fixed by the remaining conformal symmetries to be:\footnote{
We use $x^\mu$ to denote ambient space coordinates and $y^i$ to denote coordinates
along the defect. The coordinates in the transverse plane are $w^\alpha$ 
such that $x\rightarrow (w, y)$ and we will often use complex coordinates $w,\bar{w}$ in this plane. $x^M$ will denote gravitational bulk coordinates.  } 
\be
\left< D_w (y) D_{\bar{w}} (y') \right> = \frac{ C_D}{ 2 |y-y'|^{2(d-1)} }
\ee
where our conventions are such that correlation functions implicitly include the defect operator $\tau_\Sigma$ (unless otherwise stated) and are normalized such that $\left< 1 \right> = 1$. Here $C_D$ is an unfixed parameter
that depends on the particular CFT under consideration.  One of the goals of this paper is to calculate $C_D$ in a holographic theory. 

We now examine the correlation function $\left< T^{\mu\nu}(x) D_\alpha(y') \right>$ between
an ambient space stress tensor and the displacement operator.  This 
is similarly fixed by the remaining conformal symmetries up to three parameters.
Using the Ward identity \eqref{wi}, and the fact that an integrated displacement
operator against a uniform vector field $\delta w^\alpha$ simply translates the defect, these three parameters can be related to the two parameters $C_D$ and the twist operator dimension $h_\Sigma$. The later being canonically defined via the one point function of the ambient stress tensor in the presence of the defect $\left< T^{ij} \right> = - \frac{h_\Sigma}{2\pi }\delta^{ij}/|w|^d$
which we quote only for the parallel components of the stress tensor. 

We choose to represent $\left< T^{\mu\nu}(x) D_\alpha(y') \right>$ in a way which is convenient
for comparison to our holographic analysis in the next section. Firstly we exploit the symmetries
of this correlator to send the displacement operator to $\infty$ along the defect: 
$D_w(\infty) = {\rm lim}_{y' \rightarrow \infty} |y'|^{2(d-1)} D_w(y')$, without loss of generality.  
We then Weyl rescale the ambient metric so that we work in Hyperbolic coordinates $~ \mathcal{H} = \mathbb{S}^1 \times \mathbb{H}_{d-1}$  appropriate for the hyperbolic black hole description
of the twist operator. In the orbifold CFT the radius of the $\mathbb{S}^1$ factor is $2\pi n$ and the $\mathbb{Z}_n$ symmetry rotates this factor by $2\pi$.  The $\mathbb{H}_{d-1}$ factor has unit curvature radius. The metric is:
\be
ds^2_\mathcal{H} = d\tau^2 + \frac{d \rho^2 + \delta_{ij} d y^i dy^j}{\rho^2} 
= e^{2 \Omega} ds^2_{\mathbb{R}^d}\,
\quad e^{\Omega} = \frac{1}{\rho}
\ee
where the transverse coordinates are $w = \rho e^{ i \tau}$.
In these Weyl related coordinates the twist defect $\Sigma$ now lives at the conformal boundary of $\mathbb{H}_{d-1}$.
We should  rescale the ambient operators of dimension $\Delta$ as $\mathcal{O}_\mathcal{H} = e^{-\Delta \Omega} \mathcal{O}_{\mathbb{R}^d} $ , but leave alone the operators located on $\Sigma$.

Without further ado the stress tensor displacement correlator is:
\begin{align} 
\label{TD}
& \left< T_{\mu \nu}(x) D_w(\infty) \right>_{\mathcal{H}}  d x^\mu dx^\nu 
=  \\ &  \frac{\rho^{d-2} \bar{w} }{2\pi} \left(   \left(  C_D -   S_d h_\Sigma \right) \frac{d\rho^2 - \frac{1}{d-2} d \vec{y}^2 }{\rho^2} 
+ \frac{S_d h_\Sigma}{2} \frac{d \bar{w}^2  }{\bar{w}^2} \right) \nonumber
\end{align}
where $S_d = d \left(4/\pi\right)^{\frac{d-1}{2}} \Gamma( (d+1)/2)$.  A similar result holds
for the $\bar{w}$ component of the displacement operator where we switch $w \leftrightarrow \bar{w}$. 

\vspace{8pt}
\noindent \emph{Holographic description of displacement operator.} We will now try to reproduce this from a gravity calculation and in
so doing fix $C_D$. The goal will be to find a linearized
metric fluctuations about the hyperbolic black with the appropriate symmetries
and whose holographic stress tensor \cite{Balasubramanian:1999re} agrees
with \eqref{TD}. This metric fluctuation will have the interpretation as a bulk to boundary
propagator for the displacement operator. Ultimately we would like to work in the absence of a CFT metric deformation so we expect that such a bulk metric fluctuation will look normalizable in the sense that an appropriate Fefferman-Graham expansion of the fluctuation has zero deformation to the boundary metric. Note the fluctuation will only be superficially normalizable since there will be a displacement operator insertion but this is hidden at $y' \rightarrow \infty$ in our coordinate system.

The metric dual to the undeformed twist operator and which solves Einstein's equations is \cite{emparan1999ads,hung2011holographic}:
\be
\label{hbh}
ds^2 = f d\tau^2 + \frac{dr^2}{f} + r^2  \frac{d \rho^2 + \delta_{ij} d y^i dy^j}{\rho^2} 
\ee
where $f = r^2 - 1 - M r^{2-d}$. The mass $M$  determines the replica index of the twist
operator $n$ by solving the equations $f(r_H)=0$ and $n = 1/(2f'(r_H))$. Here $r_H$ is
the horizon radius and $2\pi n$ is the inverse Hawking temperature of the horizon. The zero
mass $M=0$ case corresponds to $n=1$ and is secretly the $AdS_{d+1}$ metric
written in funny coordinates. 

For the CFT living on the hyperbolic coordinates $\mathcal{H}$ 
the symmetry group $SO(2)\times SO(d-1,1)$ is manifest as the isometry group of  $\mathcal{H}$ which extends trivially into the bulk for the metric \eqref{hbh}. Thus it is natural to expect that the sought after bulk metric fluctuation respect the scaling and rotation symmetry of the displacement operator insertion on the boundary. This, along with the form of the stress tensor in \eqref{TD}, suggests the following fluctuation ansatz
in radial gauge:
\begin{align}
\nonumber
\hspace{-.132cm} h_{MN}^{D_w(\infty)} & dx^M dx^N  =\rho^{d-2} \bar{w} \left( 
f \left[ \mathbf{k}^\tau_\tau d \tau^2  +  i  \mathbf{k}^\tau_\rho  d \tau \frac{d \rho}{\rho} \right] \right.\\ & \left.
\qquad + \frac{r^2}{\rho^2}\left[ \mathbf{k}^\rho_\rho d \rho^2 +   \mathbf{k}^y_y
\delta_{ij} d y^i d y^j \right] \right)
\label{fluct}
\end{align}
where the $\mathbf{k}^M_N$ are functions only of the radial direction $r$.
The ansatz in \eqref{fluct} has definite scaling under $(\rho,y^i) \rightarrow \lambda (\rho, y^i)$ and
under rotations $\tau \rightarrow \tau + c$ and is independent of $y_i$. 

Indeed it is easy to find a very simple solution to Eintein's equations for $\mathbf{k}$:
\be
\label{asoln}
 \mathbf{k}^\rho_\rho = -(d-2) \mathbf{k}^y_y   =  k(r) \qquad \mathbf{k}^\tau_\rho = 0 
\qquad \mathbf{k}^\tau_\tau = 0 
\ee
where $k(r)$ satisfies the following differential equation:
\be
k''(r) + \left( \frac{d-1}{r} + \frac{f'}{f} \right) k'(r) - \frac{r^2 + (d-3)f}{r^2 f^2} k(r) = 0.
\label{eq:diffeqk}
\ee
This equation was first written down for the case $d=4$ in \cite{dong2016shape} where it was used to find certain $ln$ divergences that appear in the shape dependence of Renyi entropies in even dimensions. The coefficients of these divergences are in turn related to $C_D$ \cite{Bianchi:2015liz}.
 While our methods look naively very different to \cite{dong2016shape} they yield the same radial equation and the same prediction for $C_D$ when $d=4$. We will make further contact with \cite{dong2016shape} soon. For now we note that our method allows for simple generalization to arbitrary dimensions without relying on the $ln$ divergences in entanglement that only appear in even dimensions.

This equation can be solved numerically and up to normalization is fixed by the requirement
of regularity at the horizon $r=r_H$. From this solution we have the connection problem:
\begin{align}
\nonumber
k(r) & \sim \alpha_H (r-r_h)^{n/2} + 0 \times  (r-r_h)^{-n/2} \ldots \\
k(r) & \sim \alpha \frac{\sqrt{r^2-1}}{r}+ \beta r^{-d} + \ldots  \label{connect}
\end{align}
between the horizon and the boundary respectively. This uniquely defines the ratio $\beta/\alpha$
that will determine $C_D$ as we will see shortly. Note there is no reason for $\alpha$ to be zero
so naively the solution we wrote in \eqref{asoln} is \emph{not} normalizable. 

This is easily remedied. The radial gauge leaves unfixed, to first order in the fluctuations, three large diffeomorphisms determined in terms of three constant parameters. On the boundary
CFT these act as a diffeomorphism in the transverse plane to the defect as well as a Weyl rescaling
of the boundary metric. We leave their form to Appendix~\ref{app:diffs}, where we further elucidate
their meaning from a purely CFT perspective. The three parameters are sufficient to remove
the boundary metric deformation and at the end of the day we find
the following ``normalizable'' solution to Einstein's equations:\footnote{One might wonder why this is possible at all. On the boundary there are four components of the metric that need to be tuned
to or kept zero, however we only have 3 parameters in the diffeomorphism. So generically this is not possible. However the condition for a defect operator to appear in the ambient operator to defect OPE in holographic theories is exactly this normalizability \cite{Aharony:2003qf} and since we know the displacement
operator of dimension $d-1$ must appear in this OPE this had to work out. Had we started with
a different ansatz in \eqref{fluct} with the replacement out front of $\rho^{d-2} \rightarrow \rho^{\widehat{\Delta}-1}$ 
we would only find normalizable solutions for specific discrete values of $\widehat{\Delta}$. A more systematic study of these other defect operators will be left to future work. }
\begin{align}
\label{fullsoln}
\mathbf{k}^\rho_\rho
&= k(r) + \hat{\alpha} \left(  \frac{\sqrt{f}}{r} -(d-1)^2 Z^{\rho} \right)  \\
\mathbf{k}^y_y
\nonumber
& = -\frac{k(r)}{d-2} + \hat{\alpha} \left(  \frac{\sqrt{f}}{r} + (d-1)Z^{\rho}  \right)   \\
\nonumber
\mathbf{k}^\tau_\tau
&= \hat{\alpha} \left( \frac{f'}{2\sqrt{f}} + Z^{\tau}  \right)   \quad
\mathbf{k}^\tau_\rho
=  (d-1) \hat{\alpha} \left(  \frac{r^2}{f} Z^{\rho} + Z^{\tau} \right)  
\end{align}
where $\hat{\alpha} = \alpha/d/(d-2)$ and
\be
Z^{\rho} =  1 + \int^r_\infty \frac{dr'}{r'^2 f(r')^{1/2}} \quad 
Z^{\tau}  = -1+ \int_\infty^r \frac{ dr'}{f(r')^{3/2}}
\ee
We claim that this solution \eqref{fullsoln} to the linearized Einstein's equations should be thought
of as a bulk to boundary propagator for the displacement operator. Note however we have not specified
the overall normalization $\alpha$ of this solution - we will do this shortly. 

Using this result we can now calculate the desired displacement operator to ambient stress tensor two point function simply by reading of the holographic stress tensor from the near boundary
expansion of the solution \eqref{fullsoln}. \footnote{We must remember to multiply the resulting holographic stress tensor by a factor of $n$ coming from the sum of stress tensors on
each copy of the original CFT in the tensor product.  This constructs the orbifold stress tensor relevant for the dCFT correlation functions. }
We find exactly the desired form \eqref{TD} allowing us to fix both $\alpha$ and $\beta$ in terms of $C_D$ and $h_\Sigma$. (See Appendix~\ref{app:diffs} (\ref{fixa}-\ref{fixb}).)
Similarly the one point function of the stress tensor can be calculated from the background metric
\cite{hung2011holographic}, $h_\Sigma = - n M/(8G_N)$. This allows us to fix the normalization $\alpha = S_d (d-2)/(d-1)$.
We will confirm this particular normalization in the next section 
by showing that, $\int \delta w \times$~\emph{this~soution}, can be identified
with a small displacement of the gravitational dual of the defect operator. 
The final expression for
\be
C_D = \frac{n  S_d d}{16G_N (d-1) } \left(  2 (d-2) \frac{\beta}{\alpha} - M \right)
\ee
is  determined in terms of the ratio $\beta/\alpha$ which is fixed by the numerical connection problem specified in \eqref{connect}.  We make some plots of $C_D$ as a function of $n$ in various dimensions in Figure~\ref{fixcd}. 

\myfig{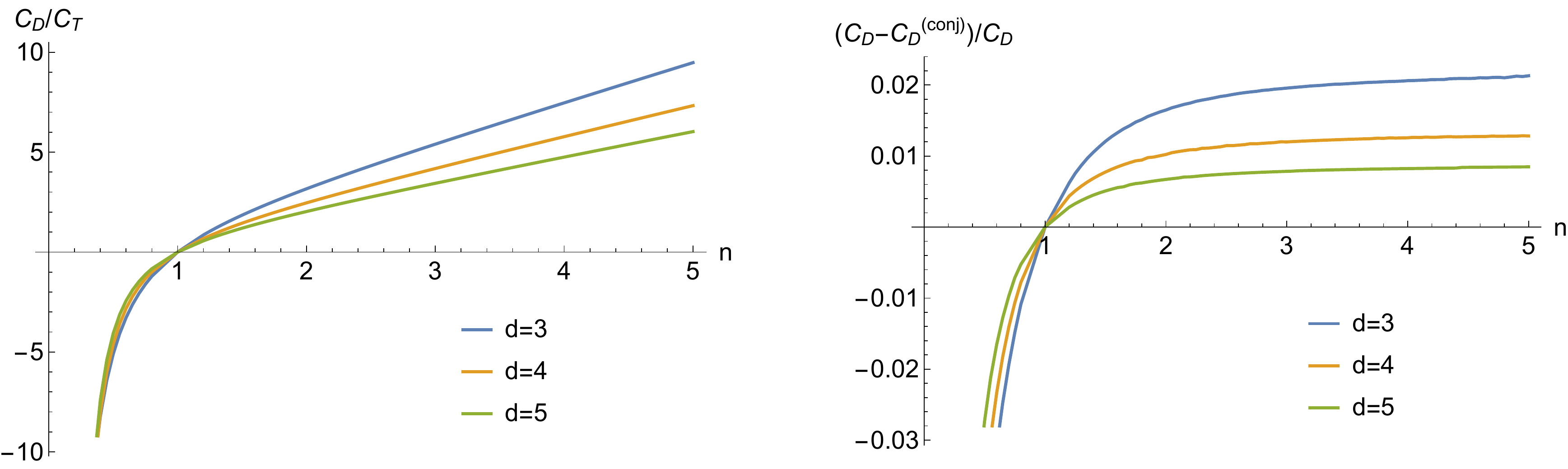}{8.7}{\small{\textsf{Plots showing $C_D$ as a function of $n$ in holographic
theories with an Einstein gravitational dual. In the \emph{left} plot we have normalized
to the coefficient of the stress tensor two point function $C_T$ and in the \emph{right} plot we have shown the relative error of the holographic answer from the conjecture of \cite{Bianchi:2015liz} which
used results from free theories, and other conjectures, to guess that $C_D^{conj} = S_d h_\Sigma$ in the conventions used here. }} \label{fixcd}}

If the displacement operator
is not inserted at $\infty$ we can use a simple inversion transformation on Hyperbolic coordinates: $(\rho, y^i) \rightarrow (\rho,y^i)/(\rho^2 + |y-y'|^2)$, 
to find the bulk to boundary propagator for $D_w(y')$. To simplify the discussion, we first consider
the fluctuation in the form \eqref{asoln} without applying the bulk diffeomorphism of \eqref{fullsoln}. 
This should be sufficient for calculating most correlators involving displacement operators
up to the action of the appropriate boundary diffeomorphism plus Weyl transformation on the operators in lower point functions.  
In this case the full bulk to boundary propagator may be written as:
\begin{align}
&h_{mn}^{D_w(y')}   \label{btob} \\
&=  \hat{k}(r) r^2  \frac{\bar{w}}{\rho} \left( \hat{\nabla}_m \hat{\nabla}_n - \hat{g}_{mn} \right)\left[ \frac{ c_{d-1} \rho^{d-1}}{( \rho^2 + (y-y')^2)^{d-1}}\right] \nonumber 
\end{align}
where $\hat{k}(r)$ is the radial solution to \eqref{eq:diffeqk} normalized so that $\alpha \rightarrow 1$ in \eqref{connect}
and $c_{d-1} =  \frac{S_d}{d(d-1)}$. The hatted covariant derivatives and metric are for the factor $\mathbb{H}_{d-1}$ where we use indices $m,n \ldots$ for the coordinates $(\rho,y^i)$ on this factor. 
We recognize the object in square brackets in \eqref{btob} as the bulk to boundary propagator for a scalar field $\phi^\alpha$ on $\mathbb{H}_{d-1}$ ``dual''  to a dimension $d-1$ operator living on the boundary $\Sigma$ of $\mathbb{H}_{d-1}$. So after integrating the displacement operator against the  vector $\delta w^\alpha$  we can write the most general solution as:
\begin{align}
h_{mn} & = \int d^{d-2} y' h_{mn}^{D_\alpha(y')} \delta w^\alpha(y')  \\
& = 2 \hat{k}(r) r^2  \frac{w_\alpha}{\rho} \left( \hat{\nabla}_m \hat{\nabla}_n - \hat{g}_{mn} \right) \phi^\alpha
\end{align}
where $\phi^\alpha(\rho,y)$ solves $(\hat{\nabla}^2 - (d-1) ) \phi^\alpha =0$ subject to the boundary conditions as $\rho \rightarrow 0$:
\be
\label{bcphi}
\phi^\alpha(\rho,y) \rightarrow \rho^{-1} \delta w^\alpha(y) + \ldots
\ee 
Note that the near $\mathbb{H}_{d-1}$ boundary expansion of the metric looks like
\begin{align} \nonumber
h_{mn} dx^m dx^n&  \rightarrow \frac{2 \hat{k}(r) r^2}{\rho^2} \left(  w_\alpha (K^\alpha_{ij} - \frac{ {\rm tr} K^\alpha }{d-2}  \delta^{ij} ) dy^i d y^j   \right. \\ & \quad \left. + \frac{2}{d-2} w_\alpha \partial_i {\rm tr} K^\alpha d y^i d\rho + \ldots \right)  
\label{simpsoln}
\end{align}
where $K^\alpha_{ij} = \partial_i \partial_j \delta w^\alpha$ is the extrinsic curvature of the boundary defect.  This corresponds to the gravity solution dual to the so-called deformed hyperbeloid  \cite{dong2016shape} which was studied in a near $\mathbb{H}_{d-1}$ boundary expansion as above. Here we have extended the results of \cite{dong2016shape} to a full solution of linearized Einstein's equations.  The boundary metric extracted from the $r\rightarrow \infty$ limit corresponds to the deformed defect in a Gaussian normal-like coordinate system. In particular the action of the bulk diffeormophism which moves us from the solution in \eqref{fullsoln} to that of \eqref{asoln} has the effect of adapting  the deformed defect to this aforementioned coordinate system as discussed further in Appendix~\ref{app:diffs}. 

Note the form of the metric plus fluctuations at the horizon as $r \rightarrow r_H$ is:
\begin{align}
\label{lm}
ds^2 &= d R^2 + \frac{ R^2 d\tau^2}{n^2}  + \ldots \\ &+ \left( \gamma_{mn} + R^n e^{ - i \tau} \hat{K}_{mn}^w
+R^n e^{  i \tau} \hat{K}_{mn}^{\bar w} + \ldots \right) d x^m d x^n  \nonumber
\end{align}
where $ R \propto \sqrt{r-r_H}$, $\gamma_{mn} \propto \hat{g}_{mn}$ and the $\hat{K}_{mn}^\alpha \propto \left( \hat{\nabla}_m \hat{\nabla}_n - \hat{g}_{mn} \right) \phi^\alpha$ are traceless by virtue of the equation of motion for $\phi^\alpha$.
This is exactly the near horizon expansion employed by Lewkowycz-Maldacena \cite{lewkowycz2013generalized} for studying the replica trick in holographic theories. 

Here we have focused on sourcing the displacement operator and the bulk to boundary propagator. The expectation value of $D_\alpha$ in the presence of other sources or in excited states
can
be read off from the asymptotic value:
\be
\phi^\alpha \rightarrow \frac{c_{d-1}}{C_D} \rho^{d-1}  \left< D^\alpha(y) \right>
\ee
which we must extract by taking the $r \rightarrow \infty$ limit first on the bulk metric fluctuation:
\be
\label{read}
\frac{d}{16 \pi G_N}  \lim_{\rho \rightarrow 0} \lim_{r \rightarrow \infty} \left( \frac{ r}{\rho} \right)^{d-2} i \oint d w \delta^{ij} h_{ij}  = b_D \left< D_w \right>
\ee 
where $b_D =  \frac{ (d-2) }{(d-1)} \frac{(C_D - S_d h_\Sigma)}{C_D} $
This is the gravitational dual of the defect OPE (dOPE!) discussed in \cite{Aharony:2003qf}. To arrive at \eqref{read} above
we used the normalizable form of  solution \eqref{simpsoln} which for completeness we give in the Appendix~\ref{app:diffs} \eqref{fullfinalsoln}.

\vspace{8pt}
\noindent \emph{Entanglement entropy limit}.
As usual the limit $n \rightarrow 1$ is of interest since  this is how we extract the entanglement entropy. Several
things simplify in this limit, essentially because the original Hyperbolic black hole, about which we are perturbing, turns into $AdS_{d+1}$. Of course analytically continuing in $n$
for a general twist defect operator is a tricky business, however in holographic theories it is known how to proceed \cite{lewkowycz2013generalized} and we have implicitly adopted this procedure by virtue of \eqref{lm}. Firstly lets focus on the displacement operator two point function. We can use standard methods to study the ODE in a perturbative series to extract:
\be
C_D \approx S_d h_\Sigma \left( 1 + \frac{(d-2)}{2d (d-1)^2} (n-1) + \ldots \right)
\ee
where $S_d h_\Sigma \rightarrow (n-1)  2\pi^2 C_T/(d+1)$ for $C_T$ the coefficient of the (non defect) stress tensor two point function. We have written the answer as a departure from the conjectured relation in \cite{Bianchi:2015liz}. It was proven that relation is true in the limit $n \rightarrow 1$ \cite{Faulkner:2015csl} however it was shown to fail in \cite{dong2016shape}
for holographic theories in $d=4$ and we confirm this in arbitrary dimensions.

Now consider the full gravitational solution for the displacement operator insertion. In the limit $n\rightarrow 1$ the solution for $\hat{k}(r)$ becomes:
\be
\label{kpert}
\hat{k}(r) =   \frac{ \sqrt{r^2-1} }{r}\left( 1  - \frac{(n-1)}{d-1} \int^r_\infty \frac{ dr' r'^{1-d} (r'^2+1) }{(r'^2-1)^2} + \ldots \right)
\ee
where we have matched to the regular horizon behavior in \eqref{connect}. We can now apply a bulk coordinate transformation (the details of which we leave to Appendix~\ref{app:null}) which takes us out of radial gauge, but puts this solution in the incredibly simple form:
\be
\label{nullsoln}
\left(\delta d s^2\right)^{D_w(\infty)}  = (n-1) c_{d-1} \left(\frac{\rho}{r}\right)^{d-2} \frac{(d u^-)^2}{u^-} + \mathcal{O}(n-1)^2
\ee
where $u^{\pm} = \sqrt{r^2-1} \rho e^{\pm i\tau}/r$. Actually we recognize these new coordinates
along with $z = \rho/r$ as those of $AdS_{d+1}$ in the Poincare patch: $ds^2 = \frac{ du^+ du^- + \delta_{ij} d y^i dy^j + z^2}{z^2}$, 
which is appropriate for the $n \rightarrow 1$ limit. In these new coordinates  \eqref{nullsoln} turns out to be, up to a factor of $(n-1)$,  the bulk metric response to the insertion of a null energy operator in real times:
\be
\mathcal{E}_+(y') = 2\pi \int_0^\infty d u^+ T_{++}(u^+,u^-=0, y')
\ee
Examining the bulk to boundary propagator for the null components of the metric response (see for example \cite{Liu:1998bu}), analytically continuing it to real time and performing the null integral, we have:
\be
\left( \delta g_{--} \right)^{\mathcal{E_+}(y')} =  \frac{ c_{d-1} z^{d-2}( (y-y')^2 + z^2)^2}{ u^- \left(  u^+ u^- +  (y-y')^2 + z^2\right)^{d+1} }
\ee
The limit $y' \rightarrow \infty$ (after multiplying by $|y'|^{2(d-1)}$) then agrees with \eqref{nullsoln}. 
This is consistent with the recent discussion of the shape dependence of modular Hamiltonians \cite{Faulkner:2016mzt} and entanglement entropy \cite{edensity,Faulkner:2015csl} where these null energy operators come about from a conformal perturbation theory calculation in the boundary CFT \cite{Faulkner:2014jva}.

From this result one might be tempted then to simply identify $\lim_{n \rightarrow 1} \frac{1}{n-1} D_w
=  \mathcal{E}_+$? But this is not always correct since the analytic continuation in $n$ depends on the exact correlator under consideration. For example this identification would tell us that the entanglement density naively vanishes as $(n-1)$. The  entanglement density is defined via the second variation of the entanglement entropy $\delta^2 S_{EE} \sim - \int d y_1 \delta w^\alpha \int d y_2 \delta w_\alpha \mu(y_1,y_2)$. To extract this from either the displacement operator or the null energy operator we should compute:
\begin{align}
\mu(y_1, y_2) &
\sim \lim_{n\rightarrow 1} \frac{1}{n-1} i \oint d\bar{w} \left< T_{\bar{w}\bar{w}}(y_1,w) D_{w}(y_2) \right> \\ 
& \hspace{-1.5cm}  \sim  \lim_{\mathcal{V} \rightarrow \infty} \frac{2\pi}{\mathcal{V}} \int_{e^{-\mathcal{V}/2}}^{e^{\mathcal{V}/2}} \hspace{-.4cm} d u^- \left< T_{--}(y_1, u^-, u^+=0) \mathcal{E}_+(y_2) \right>_{\text{\tiny{CFT}}}
\end{align}
where $\mathcal{V}$ is a divergent volume factor (in Rindler space) that needs to be removed when one does the integral over null coordinates.  The later correlation function is evaluated in the CFT with no mention of the defect. 
This later relation is derived in depth in a forthcoming paper \cite{edensity}. The former relation is a  consequence of the ward identity \eqref{wi} and the form of the correlator in the limit $n \rightarrow 1$. These are clearly two different computations however they look remarkably similar - the later being the real time version of the former. In particular the later
does not require any careful $n$-analytic continuation that is mostly only available for theories with a gravitational dual. 

\vspace{8pt}
\noindent \emph{Extensions.}   Given the bulk to boundary propagator studied in this paper, it is possible to compute more complicated correlation functions involving displacement operators. The answers take the form of Witten diagrams and we give a few examples in Figure~\eqref{witt}. We will not attempt to compute any of these diagrams leaving this to future work, however we mention  a few potential applications in the caption.

\myfig{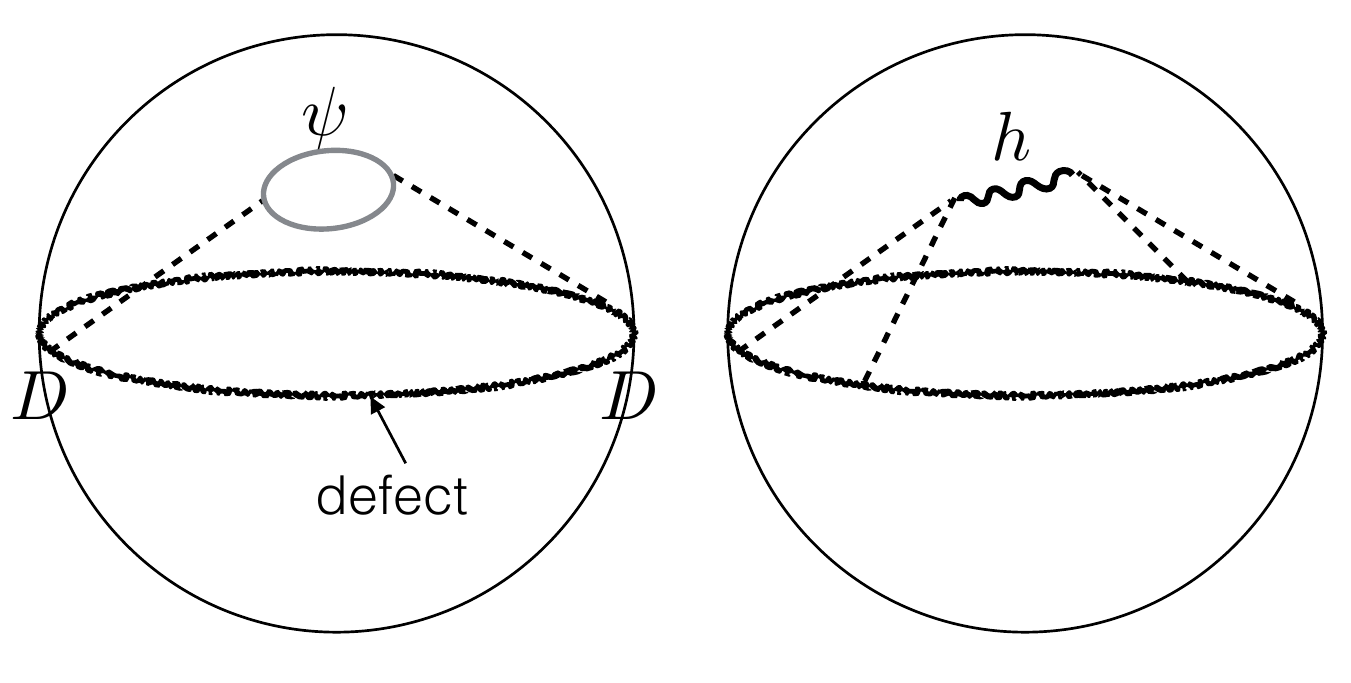}{8}{ \label{witt} \small{\textsf{
Witten diagrams \cite{Witten:1998qj} for the displacement operator. The ball represents the Euclidean hyperbolic black hole and for clarity we have shown 1 co-dimension to the defect.
The loop diagram on the left would compute the perturbative $1/N$ correction to $C_D$. 
The four point function of displacement operators is the next correction to the expansion
of the shape dependence of the Renyi entropies about the spherical or planar cut:
$
S_n( \Sigma) \sim S_n(\Sigma_0 ) +\sum_N (N!)^{-1}  \int \delta w^{\alpha_1} \ldots \int \delta w^{\alpha_N} \left< D_{\alpha_1}  \ldots D_{\alpha_N} \right>
$ 
}}}

The study of displacement operators on defects has consequences in situations other than for Renyi entropies. Defects can arise in critical phase of matter and their dynamics is controlled via the displacement operator correlation functions \cite{Dias:2013bwa}.
Thus it is of interest to extend this analysis to such holographic models.

\emph{Acknowledgments} 
It is a pleasure to thank Rob Leigh, Onkar Parrikar and Huajia Wang for discussions on related topics. 
Work supported in part by the DARPA YFA, contract D15AP00108.

\appendix

\section{Unfixed diffeomorphisms}
\label{app:diffs}

Working with the fluctuation in the form \eqref{fluct} we can make the following diffeomorphism:
\be
\label{diffform}
(r,\rho,\tau) \rightarrow (r,\rho,\tau) + \rho^{d-2} \bar{w}\left( X_r(r), \rho X_\rho(r), i X_\tau(r) \right)
\ee
These preserve the radial gauge as long as:
\be
\label{det}
X_r(r) = \delta r f^{1/2} \quad X_\tau'(r) = \frac{\delta r}{f^{3/2}} \quad X_\rho'(r)=
-  \frac{(d-1) \delta r}{r^2 f^{1/2}}
\ee
where $\delta r$ is an integration constant and there are two more constants associated 
to the differential equations in \eqref{det} such that $\lim_{r \rightarrow \infty} X_\tau(r) = \delta \tau$
and $\lim_{r \rightarrow \infty} X_\rho(r) = \delta \rho$. These act on the $\mathbf{k}$ fluctuations as:
\begin{align}
\mathbf{k}^\tau_\tau &\rightarrow \mathbf{k}^\tau_\tau  +  \left( 2 X_\tau + \frac{\delta r f'}{f^{1/2}}  \right) \\
\mathbf{k}^\tau_\rho & \rightarrow \mathbf{k}^\tau_\rho - 2 \left( \frac{r^2}{f} X_\rho -  (d-1) X_\tau \right) \\ 
\mathbf{k}^\rho_\rho & \rightarrow \mathbf{k}^\rho_\rho + 2 \left( (d-1) X_\rho + \frac{\delta r f^{1/2}}{r}  \right) \\
\mathbf{k}^y_y & \rightarrow \mathbf{k}^y_y + 2 \left( - X_\rho + \frac{\delta r f^{1/2}}{r}  \right) 
\end{align}
and picking parameters $\delta \rho = - (d-1) \delta r$ and $\delta \tau = - \delta r$ with 
$\delta r = \frac{\alpha}{2d(d-2)}$ we can use this diffeomorphism to remove the deformation of the boundary metric implied by the solution \eqref{asoln}. Only after we do this removal can we claim a gravitational solution for the displacement operator, since that involves no boundary metric deformation.

The resulting holographic stress tensor for this fluctuation takes the form in \eqref{TD} with the following identifications:
\begin{align}
\label{fixa}
C_D  - S_d h_\Sigma &= \frac{ n d}{ 8 G_N } \left( \beta + \frac{M}{2d} \alpha \right) \\
\label{fixb}
\frac{S_d h_\Sigma}{2} &= - \frac{n  (d-1)}{16 G_N (d-2)} \alpha M
\end{align}

On the boundary theory the three bulk diffeomorphisms act as two boundary diffeomorphisms and a Weyl rescaling. For the parameters identified above we have:
\begin{align}
\xi^{D_w(\infty)}  & = - \delta r \rho^{d-2} \bar{w} \left( (d-1) \rho \partial_\rho + i \partial_\tau \right) \\
\delta \Omega^{D_w(\infty)} &=  \delta r \rho^{d-2} \bar{w}
\end{align}
where we are working for now with the CFT living on the hyperbolic space $\mathcal{H} = \mathbb{S}^1 \times \mathbb{H}_{d-1}$.  We can understand the meaning of these in various ways. To begin with we note that the stress tensor displacement operator correlation function can be written as:
\begin{align}\nonumber
&\left< T_{\mu\nu} D_\alpha(\infty) \right>_{\mathcal{H}} =\left<  \left( \left[ \mathcal{L}_{\xi^{D_\alpha(\infty)}} + \delta^W_{\delta \Omega^{D_\alpha(\infty)}} \right] T\right)_{\mu\nu} \right>_{\mathcal{H}} + Q_{\mu\nu\alpha} \\
&Q_{\mu\nu \alpha} dx^\mu d x^\nu 
\label{aQ} \\
& = \frac{\rho^{d-2} w_\alpha}{\pi} \left( 2 C_D - S_d h_\Sigma \frac{3d-4}{d-1} \right)  \frac{d\rho^2 - \frac{1}{d-2} d \vec{y}^2 }{\rho^2}  \nonumber
\end{align}
where $\mathcal{L}$ is a Lie derivative and $\delta^W$ is an infinitesimal Weyl rescaling with the appropriate weight ($d-2$ in the case of $T_{\mu\nu}$.) Here the stress tensor one point function $\left< T_{\mu\nu} \right>_{\mathcal{H}}$ in the presence of the defect appears and we should note that we are ignoring possible scale anomalies in $\left< T_{\mu\nu} \right>_{\mathcal{H}}$ by working in general dimensions $d$. 

Further the diffeomorphism plus Weyl transformation acts on the $\mathcal{H}$ metric to give:
\begin{align}
\nonumber
\left(\left[ \mathcal{L}_{\xi^{D_\alpha(\infty)}} + \delta^W_{\delta \Omega^{D_\alpha(\infty)}} \right] g\right)_{\mu\nu} & dx^\mu d x^\nu
= \\-  \frac{4 S_d (d-2)}{d-1}  \rho^{d-2}w_\alpha & \frac{ d\rho^2 - \frac{1}{d-2} d \vec{y}^2 }{\rho^2} 
\label{metdef}
\end{align}
The conclusion is that if we invert the diffeomorphism and Weyl transformation we study the twist defect in the presence of a particular metric deformation - minus that given in \eqref{metdef} -  and we expect the stress tensor response to this deformation as $Q_{\mu\nu \alpha}$ alone, as given in \eqref{aQ}. This is the meaning of the  AdS/CFT solution in \eqref{asoln}. Since this is clearly a much simpler solution to study (compared to the ``normalizable'' solution in \eqref{fullsoln}) it is useful to understand what its interpretation is in even more detail. To do this we will now examine the integrated displacement operator and the resulting diffeomorphism plus Weyl transformation:
\begin{align}
\xi &= - \int d^{d-2} y' \xi^{D_\alpha(y')} \delta w^\alpha(y')  \\
\delta \Omega &= - \int d^{d-2} y' \delta \Omega^{D_\alpha(y')} \delta w^\alpha(y') 
\end{align}

Still working in the conformal frame of $\mathcal{H}$ we find:
\begin{align}
\xi &= \rho^{-1} w_\alpha \hat{g}^{mn} \partial_m \phi^\alpha \partial_n
 + \rho^{-1} \epsilon_{\alpha \beta} \phi^\alpha w^\beta \partial_\tau \\
 \delta \Omega &= - \rho^{-1} w_\alpha \phi^\alpha
\end{align}
where $\phi^\alpha$ was defined around \eqref{bcphi} and $\hat{g}$ refers to the metric on $\mathbb{H}_{d-1}$ as in the main text. Moving back to flat space gives the clearest interpretation of this result. When we shift back to flat space the diffeomorphism is un-altered but the Weyl deformation shifts
as $\delta\Omega' = \delta \Omega - \rho^{-1} \xi_\rho$. Now we expand close to the entangling surface $\rho \rightarrow 0$ to find:
\begin{align}
\label{xiexp}
\xi \rightarrow & \quad \delta w^\alpha \partial_\alpha + \delta^{ij} w_\alpha \partial_i \delta w^\alpha \partial_j
 \\ & +\frac{1}{d-2} \left( \partial^2 \delta \bar{w} w^2 \partial_w  +\partial^2 \delta w \bar{w}^2 \partial_{\bar{w}} \right)  + \ldots \nonumber  \\
 \delta \Omega' \rightarrow & \quad - \frac{w_\alpha}{d-2} \partial^2 \delta w^\alpha + \ldots
\end{align} 
Recall that due to the integrated displacement operator the twist defect is originally located at $w^\alpha = - \delta w^\alpha$ (see \eqref{deform}). But it is now it is clear from the leading term in \eqref{xiexp} that $\xi$ exactly moves the twist defect back to the origin $w^\alpha = 0$. The result is that we are naturally encouraged to work in a coordinate system adapted to the deformed twist defect, which is something like Gaussian normal coordinates. In these new coordinates the deformed metric close to the twist defect is:
\begin{align}\nonumber
ds^2 = d w d\bar{w} +& \left( \delta_{ij} + 2 w_\alpha \left(K_{ij}^\alpha - \frac{ {\rm tr} K^\alpha}{d-2} \delta_{ij} \right) \right) dy^i dy^j \\& +  \frac{4}{d-2} w_\alpha \partial_i {\rm tr} K^\alpha d y^i d\rho + \ldots
\end{align}
This is the metric of deformed hyperbeloid that was studied in \cite{dong2016shape} using this $\rho \rightarrow 0$ limit. Note that at higher orders the coordinates are not quite Gaussian normal (and thus differ slightly from \cite{dong2016shape}), since there are $d \rho d y^i$ cross terms. There is no particular reason why Gaussian normal is the most natural. It seems that conformal invariance of the defect naturally picks the above metric. Also note that, since $\phi^\alpha$ is fully determined in the ambient space by a given deformation  $\delta w^\alpha$ on $\Sigma$, the full diffeomorphism, Weyl factor and metric deformation are fixed even far away from the entangling surface. 

For completeness we also give the full form of the displacement bulk to boundary propagator in the absence of any boundary metric deformation:
\begin{align}
\nonumber
  h_{mn} &= \frac{2 w_\alpha}{\rho} r^2 \left( ( \hat{k} - Z^\rho) \hat{\nabla}_m \hat{\nabla}_n - (\hat{k} - \frac{\sqrt{f}}{r} ) \hat{g}_{mn} \right)  \phi^\alpha \\
\nonumber
 h_{m\tau} &=  \epsilon_{\alpha \beta} \frac{ w^\beta}{\rho}  ( r^2 Z^\rho + f Z^\tau) \hat{\nabla}_m \phi^\alpha \\
 h_{\tau\tau} & = \frac{ w_\alpha}{\rho}  \left( f' \sqrt{f} + 2 f Z^\tau \right) \phi^\alpha
 \label{fullfinalsoln}
\end{align}

\section{Revealing the null energy insertion}
\label{app:null}

We write the solution to $\hat{k}$ in a perturbative series:
\be
\hat{k}(r) = \frac{f_0^{1/2}}{r} \left( 1 - \frac{ 2\delta_n}{d-1} u_1(r) + \ldots \right)
\ee
where $u_1'(r)=  r^{1-d} \frac{(r^2+1)}{2(r^2-1)^2}$ such that $\lim_{r \rightarrow \infty} u_1 = 0$. Also  we have defined $f_0 = (r^2 - 1)$ and $\delta_n = (n-1)$.
 The diffeomorphism that reveals the desired solution \eqref{nullsoln} starting from the original solution \eqref{fullsoln} is given by the form as in \eqref{diffform} but with:
\begin{align}
X_r &= \delta r f_0^{1/2} \left( 1 + 2\delta_n \left(  \frac{r^{-d}}{2d} - \frac{u_1}{d-1} \right) + \ldots \right) \\
X_\tau & = - \frac{\delta r r}{f_0^{1/2}} \left( 1 - \frac{2 \delta_n }{d-1} \left( r^{-d} \left( \frac{1}{2d} + \frac{1}{f_0} \right) + u_1\right) + \ldots  \right)   \nonumber \\
X_\rho &= -\frac{\delta r f_0^{1/2}}{r} \left( (d-1) - 2\delta_n \left( \frac{r^{-d}}{2d} + u_1 \right) + \ldots \right) \nonumber
\end{align}
This choice does not satisfy the requirement \eqref{det} that we stay in radial gauge, however this is necessary to move to the solution in \eqref{nullsoln}.
 
\bibliography{displacement}

\end{document}